# COVID-19 reproduction number estimated from SEIR model: association with people's mobility in 2020

Tatiana Petrova [1,*], Dmitry Soshnikov [2,3] and Andrey Grunin [1]

1. Faculty of Physics, Lomonosov Moscow State University, Moscow, Russia
2. Microsoft, Developer Relations, Moscow, Russia
3. Faculty of Computer Science, Higher School of Economics, Moscow, Russia

*Correspondence: Tatiana Petrova, Faculty of Physics, Lomonosov Moscow State University,
119992, Leninskie gory 1bld2, Moscow, Russia, Tel: +7 495 939 12 50, e-mail: tapetrova@physics.msu.ru

**Abstract:** This paper is an exploratory study of two epidemiological questions on a worldwide basis. How fast is the disease spreading? Are the restrictions for people (especially mobility restrictions) bring the expected effect? To answer the first question, we propose a tool for estimating the reproduction number of epidemic (the number of secondary infections - $R_t$) based on the SEIR model and compare it with an non-model $R_t$ estimation. To measure the $R_t$ of COVID-19 for different countries, real-time data on coronavirus daily cases of infections, recoveries, deaths are retrieved from the Center for Systems Science and Engineering (CSSE) at Johns Hopkins University. To assess the effectiveness of mobility restrictions for the COVID-19 pandemic in 2020, the correlations between the $R_t$ and people's mobility (based on the Apple mobility index) are presented. The correlations were considered for 12 countries and for most of them the correlations are negative. This shows a delay in the implementation of mobility restrictions – the countries imposed them in response to growth of new COVID-19 cases, rather than preventively.

**Keywords:** reproduction number; COVID-19; epidemic modelling; mobility index; infectious disease; mobility restrictions effectiveness

## 1. Introduction

Coronavirus disease (COVID-19) was first identified in late December 2019 in Wuhan (China) and during the first half of 2020 spread all over the word. On March 11th 2020 the coronavirus outbreak was labelled a pandemic by the World Health Organization [1]. By the middle of 2021, COVID-19 has spread to almost all countries, affecting more than 188 million people worldwide and causing more than 4 million deaths [2]. More than half of the world's population has experienced a lockdown with strong mobility restrictions. This was the first large-scale implementation of such measures in history.

One of the key points during an epidemic is to design appropriate mathematical models and tools for effective decision-making. To optimize the quarantine steps it is important to estimate parameters characterizing infectious disease transmission using real-time data and track the temporal changes in those values [3]. Another important point is to assess the effectiveness of implemented restrictions.

**The time-dependent reproduction number**

The time-dependent (effective) reproduction number ($R_t$) is one of the key parameters characterising evolution of an epidemic. This value is a function of time, and represents the expected number of secondary cases arising from a primary case infected at time $t$ [4]. If $R_t < 1$ the disease will decline and eventually die out. If $R_t > 1$ the disease will be transmitted between people, and an outbreak is likely to occur. Reducing the reproduction number below 1 is one of the main goals of implementing quarantine steps.

There are different methods to define the initial reproduction number $R_0$ and time-dependent reproduction number $R_t$. The numerical estimations of $R_t$ using different methods vary depending on implementation [4-10]. Some of the standard methods were applied to assume basic and time-dependent reproduction numbers for COVID-19 outbreak for different countries [11-15]. In review [16] the differences for estimations of COVID-19 reproduction number for China are shown, and the difference beetween forecasted values via different methods is explained by insufficient data and different estimation techniques.

**Mobility restrictions**

The mobility restrictions were used in countries worldwide to slow the spread of COVID-19. They included closing public transportation, stores, offices; travel restrictions; requirement to stay home, etc. The scale of these response measures was incredibly large and inccurred significant costs. Longitudinal data of epidemic spreading and people's mobility are now available. They can be used for investigating the long-term effects of quarantine steps, especially mobility restrictions.

Previous works have shown that social distancing orders have an impact on reduced mobility and case growth in India and the United States [17-21]. Oh, J., et al. [22] have demonstrated that mobility restrictions appeared to reduce the spread of COVID19 in many countries in the early stages of the pandemic wave, but in the later stages, once other mitigation measures are taken, the impact will weaken. Report [11] concludes that easing social-distancing restrictions should be considered very carefully, as small increases in contact rates are likely to risk resurgence even where COVID-19 is apparently under control. In the report [23] a consistent correlation between reductions in mobility and reductions in transmission intensity of COVID-19 was found for the first months of 2020. The authors developed a minimalist compartmental model to study the impact of mobility restrictions in Italy during the Covid-19 outbreak, based on SIOR model (here $O$ (Observed) — are individuals who present symptoms acute enough to be detected from the national healthcare system, Observed individuals switch into the $R$ (Removed)). It was shown that, while an early lockdown shifts the contagion in time, beyond a critical value of lockdown strength the epidemic tends to restart after lifting the restrictions.

To understand the longitudinal impact of mobility restrictions to pandemic spread, we estimated the correlations between the $R_t$ and people's mobility (based on the Apple mobility index) for 12 countries worldwide in 2020. For estimating $R_t$ in real-time, we developed an open-source tool, requiring only data that are commonly recorded during an outbreak (infections, recoveries, deaths, infection duration), with clear science based methodology and without limitations on epidemic phase. The SEIR-based reproduction number is complemented with the non-model estimation of $R_t$ to compare the calculated results. The tool also allows to compare the dynamics of $R_t$ and peoples mobility, and assess the correlation between those. The $R_t$ calculation in connection with people's mobility allows to forecast the progression of disease outbreaks and understand the effect of restrictions on epidemics spread.

## 2. Materials and Methods

To estimate the time-dependent reproduction number, the following procedure based on the SEIR model is used. We begin at each country's 1000th case of infection; then we move a sliding 7 days window over the whole period, and for each step we estimate the time-dependent reproduction number by minimizing square difference between real and predicted (by SEIR model) number of new infected cases. To explore the time-dependent reproduction number in connection with peoples mobility, we calculate the correlation between $R_t$ and peoples mobility, $R_t$ and derivative of peoples mobility.

*Data Sources*

Real-time data on COVID-19 daily cases of infections, recoveries, deaths, as well as each country's population are retrieved from the COVID-19 Data Repository by the Center for Systems Science and Engineering (CSSE) at Johns Hopkins University [24]. The data were reported by each country's official surveillance system starting January 22nd, 2020.

Peoples mobility data is sourced from Apple Mobility Index [25]. It is generated by counting the number of requests made to Apple Maps for directions. Three data streams are used: 'driving', 'walking' and 'transit' mobility for Apple device users. The data is available from January 13th (except May, 11th-12th) and is updated every day. The mobility data for China is temporarily not available.

*Data processing*

The online data of infected, recovered, deceased cases, each country's population and transposed dates are joined in one table, which gives us each country's dataframe. The beginning is each country's 1000th case of infection, then the numbers of observed cases are smoothed out by computing 3-days rolling average on all the columns. To estimate the number of expected cases, the data on infected shifted by the median incubation period of COVID-16 (5 days [26]) is used. In addition, we applied the results of Oran D.P., Topol E.J. [27], according to which 42 to 45 percent of coronavirus cases are asymptomatic.

We automatically collect all available online Apple mobility data, obtain the mean values of three data streams ('driving', 'walking' and 'transit') for each country and apply 7-days running average to reduce weekly fluctuations.

*SEIR modelling*

The SEIR model is a modification of the classical SIR approach to epidemic simulation presented by Kermack-McKendrick [28] for the number of people infected with a contagious illness in a closed population over time. The assumptions of the model are: constant (closed) population size ($N$); constant rates (e.g., transmission, removal rates - the probability of disease transmission in a single contact multiplied by the average number of contacts per person); no demography (i.e., births and deaths); well-mixed population (where any infected individual has a probability of contacting any susceptible individual that is reasonably well approximated by the average).

The SEIR model describes the connection between $S$ (susceptible - number of people who have the potential to be infected), $I$ (infected - number of infected people), $R$ (removed - number of people who are non susceptible to infection, this includes the number of deceased people as well) and $E$ (exposed - number of people who have been infected but does not show symptoms yet: it can be called a latent phase) with the following system of differential equations:

$$dS/dt = -\beta SI/N,$$
$$dE/dt = \beta SI/N - kE,$$
$$dI/dt = kE - \gamma I,$$
$$dR/dt = \gamma I.$$
(1)

$\beta$ is known as the effective contact rate. We assume that in a unit time each infected individual will come into contact with $\beta N$ people. From those people, the proportion of susceptible people is $S/N$, thus the speed at which new infections occur is considered to be $-\beta SI/N$.

$\gamma$ is the removal rate, and the number $1/\gamma$ defines the number of days during which a person stays infected. Thus the term $\gamma I$ defines the speed at which infected individuals are moved from being infected to recovered (or deceased).

$k$ is the progression rate from exposed (latent) to infected and governs the lag between having undergone an infectious contact and showing symptoms. In the equations, it brings people from the $E$ category to the $I$ category.

The rates are supposed to be constant. In the assumption that the population is completely susceptible ($S=N$), no demography (i.e., births and deaths) and the population is well-mixed, the basic reproduction number $R_0 \approx \beta / \gamma$ [29].

*Model parameters*

Model predictions based on system (1) depend on the parameters ($\beta$, $\gamma$, $k$). Optimization of the model with respect to the parameters and fitting it to the real data allows us to find the parameters that correspond to the actual outbreak. It is more consistent to optimize for $\beta$, and set the $\gamma$ and $k$ parameters according to medical studies. According to WHO [30], the median time from onset to clinical recovery for mild cases is approximately 2 weeks, and it is 3-6 weeks for patients with severe or critical disease. Among patients who have died, the time from symptom onset to outcome ranges from 2 to 8 weeks. Since the time period is wide enough, we considered the mean recovery time of 30 days, and fixed $\gamma$ as 1/30, respectively. We estimated the number of expected cases as the data on numbers of infected, shifted by the median incubation period of COVID-16 (5 days), so $k$ is fixed as 1/5.

To smooth inaccuracies in the statistics at beginning of the outbreak the calculations start from day $t_0$, where $t_0$ is the first day when the number of infected people is above 1000:

$$t_0 = \min\{t \mid I(t) > 1000\}.$$

For each country, the following denotations for the data are used: $V(t)$ - the number of total (accumulated) infected cases at day $t$, $t > t_0$, $E^{eff}_\beta(t)$, - number of exposed cases at day $t$, $G(t)$ - number of recovered cases at day $t$, $D(t)$ - number of fatal cases at day $t$. According to the results [27], from 42 to 45 percent of coronavirus cases are asymptomatic, so we use the number of asymptomatic cases as 43 percent and calculate $E^{eff}_\beta(t)$ as

$$E^{eff}_\beta(t) = E_\beta(t)/(1-0.43) = I_\beta(t-5)/0.57$$

The values corresponding to each $\beta$ are denoted as $S_\beta(t)$, $E^{eff}_\beta(t)$, $I_\beta(t)$ and $R_\beta(t)$, respectively. $S_\beta(t)$, $E^{eff}_\beta(t)$, $I_\beta(t)$ and $R_\beta(t)$ are the solutions of system (1) with initial values of:

$$S(t_0) = N - V(t_0) - E^{eff}_\beta(t) - D(t_0) - G(t_0),$$
$$I(t_0) = V(t_0),$$
$$R(t_0) = D(t_0) + G(t_0).$$

Here $N$ is the population of a country being considered.

The accumulated number of total infected people computed by the model is given by:

$$I^+_\beta(t) = I_\beta(t) + R_\beta(t) + E^{eff}_\beta(t).$$

New total infected cases each day is equal to

$$I'_\beta(t) = I^+_\beta(t+1) - I^+_\beta(t).$$

The actual number of daily new infected cases is

$$V'(t) = V(t+1) - V(t).$$

The value $\beta^*$ is defined below (we will denote the number of days in consideration by $n=7$):

$$\beta^* = \mathrm{argmin}_\beta \sum (V'(t) - I_\beta'(t))^2.$$

The process of finding argmin (the minimum value along a given axis) is a complex optimization process, and at each step the numerical solution of ODE (1) is needed. Powell's method (Powell's conjugate direction method) is used for this optimization because it does not rely on gradients and is fast enough [31].

Having obtained $\beta^*$, we calculate $R_0$ as follows:

$$R_0 = \beta^* / \gamma.$$

*SEIR model for time-dependent reproduction number ($R_t$) estimation*

Similar to the basic reproduction number $R_0$ that is a characteristic of a disease, the time-dependent reproduction number $R_t$ is considered. The $R_t$ dynamics during the pandemic takes into account isolation measures and the proportion of nonsusceptible population, and can be used to estimate the effectiveness of quarantine measures.

The same approach is followed for estimating $R_t$ in real time: start from the point at which there are 1000 infected people; move a sliding window of width $n$ (we used $n=7$) over the whole period till present day. At each point, use $n$ consecutive days to estimate $\beta$ (and, thus, $R_t$) by minimizing the square difference between the real and predicted number of new infected people.

*Non-model estimation of $R_t$*

We also calculated an estimation of $R_t$ that is not based on epidemic modelling [32, 33] and validated it using the SEIR-based model. The non-model reproduction number ($R_t^{MF}$) is one of the main indicators that the governments relied on for making decisions to restrict or weaken the lockdown. $R_t^{MF}$ is calculated by dividing the sum of the infected cases during the last 4 days by the sum infected cases in the previous 4 days:

$$R_t^{MF} = \sum_{i=5}^{8} I_{i+t-1} / \sum_{i=1}^{4} I_{i+t-1}$$

here $I_i$ is the number of infected cases for the corresponding day $i$.

*Spearman Rank Correlation*

The correlation between the $R_t$ and people's mobility (Apple Mobility Index) can be investigated with the aid of the Spearman rank correlation coefficient. It is a non-parametric correlation method for measuring the strength and degree of association between two variables. Instead of exploring a linear relationship, in the case of Spearman rank correlation, there are no strict conditions on the association, and the observed data are ranked following a specific sequence. The correlation coefficients indicate whether there is a potential relationship between $R_t$, $R_t$'s change rate, and people's mobility. The formula below can be used to calculate the Spearman rank correlation coefficient ($r_s$) is as follows:

$$r_s = 1 - \frac{6 \sum d_j^2}{n(n^2 - 1)},$$

where $d_j$ denotes the ranked difference between the $j$ pairs of variables, and $n$ denotes the number of ranks in each of two variables.

The correlation coefficient value ranges from −1 to 1. If $r_s > 0$, then there is a similar distribution and ranking of variables. If $r_s < 0$, then the variables are ranked differently. The absolute value of $r_s$ represents the degree of correlation. If the value of $r_s$ is closer to ±1, the correlation between two variables is stronger.

## 3. Results

The approach described above allows to find real-time dynamics of reproduction number and easily analyze it in connection with other factors directly affecting the epidemic spread (number of daily infected cases, peoples mobility level, rate of changes in people's mobility level). The results are presented in Figs.1-6 for 12 countries and can be extended to any country in the world reported in [19] and [20]. For each country, the time period from 1000 infected cases till December 31, 2020 was chosen. We considered only the first year of pandemic (2020) because in 2021 most of the countries started active vaccination campaign, so the model for $R_t$ estimation has been changed and also changed the connection between $R_t$ and mobility index.

*3.1. Time-dependent reproduction number for different countries*

General trends in the COVID-19 pandemic spread connected with $R_t$ dynamics in 2020 for different countries are shown in Figure 1a.

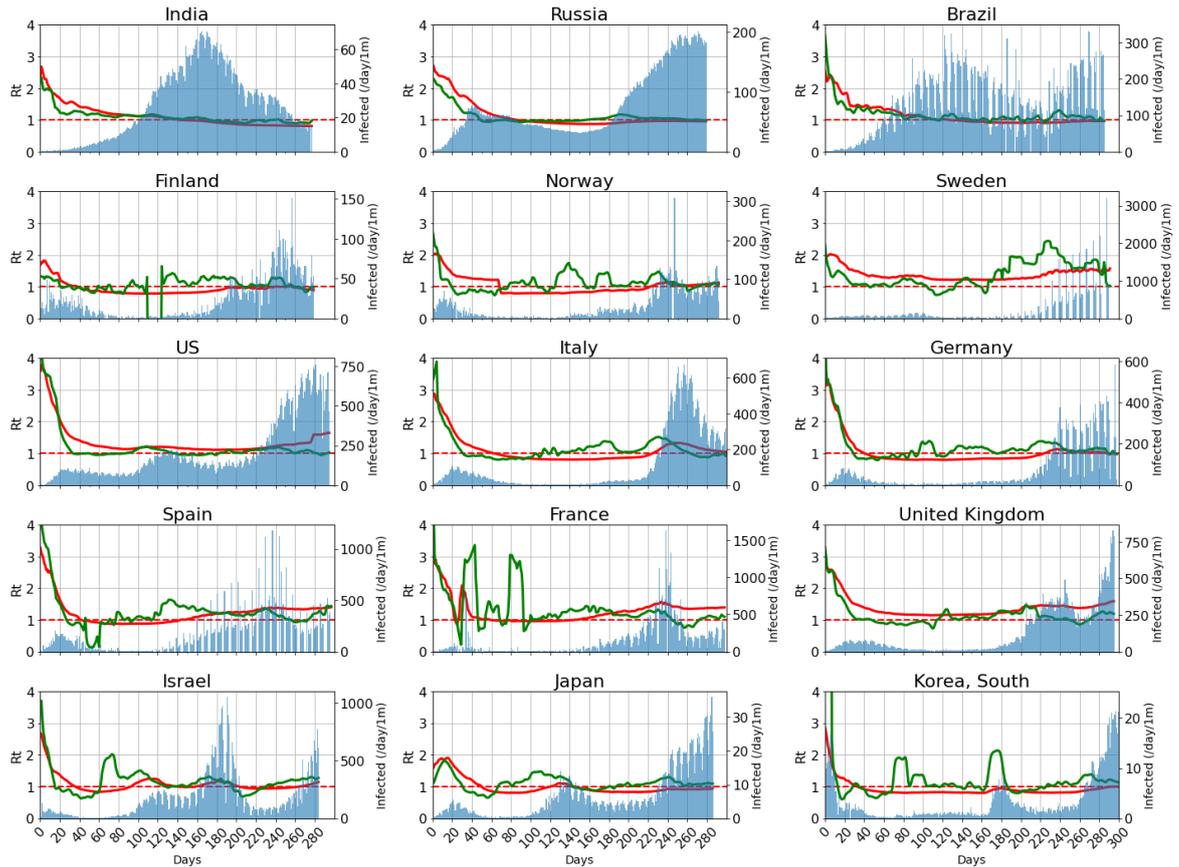

**Figure 1a.** SEIR-based time-dependent reproduction number $R_t$ (red line, left y-axis), model-free estimation of $R_t$ ($R_t^{MF}$, green line, left y-axis) and daily number of infected cases normalized by country population for each country (blue histogram, right y-axis), plotted for the period till December 31, 2020. It is shown that SEIR-based $R_t$ is more stable and less sensitive to outliers caused by inaccurate statistics, than $R_t^{MF}$. On the x-axis the number of days from the first 1000 infected cases in each country till December 31, 2020 is shown, which makes it possible to compare the epidemic spread in 2020 (before the beginning of wide vaccination campaigns). The number of new cases is measured in cases per million of people. Normalized number of new cases is shown on different scales for each graph on purpose to make the shape of the graph more visible.

Figure 1a illustrates the following main trends:
- The countries are characterized by different $R_t$ shapes — the $R_t$ curves have different initial values and different decrease rates;
- The SEIR-based $R_t$ is more smooth and less sensitive to outliers caused by inaccurate statistics than model-free estimation of $R_t$ ($R_t^{MF}$). $R_t^{MF}$ better highlights their occurrence;
- The $R_t$ values decreased significantly compared to the initial values ($R_0$) for all countries;
- One can see waves of infected cases for all of the countries, but $R_t$ numbers for these waves are much smaller than for the beginning of the epidemic and close to 1.
- For most of the countries (except Sweden, US and UK) $R_t$ during 2020 dropped below 1, but the spread of coronavirus continued.

The most informative period of the epidemic is the beginning of infection spread, and it makes sense to see the first months of the epidemic in more detail. On Figure 1b we show the graphs comparing $R_t$ and $R_t^{MF}$ for the period up to July 1, 2020 across different countries.

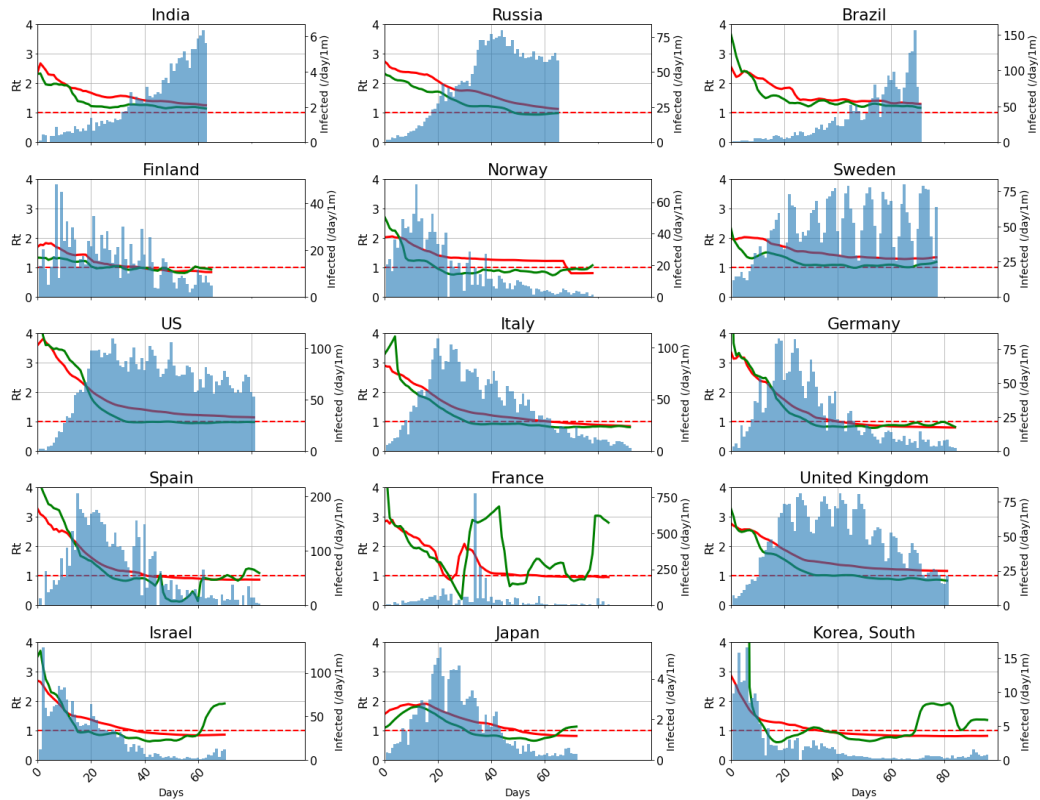

**Figure 1b.** SEIR-based time-dependent reproduction number $R_t$ (red line, left y-axis), model-free estimation of $R_t$ ($R_t^{MF}$, green line, left y-axis) and daily number of infected cases normalized by country population for each country (blue histogram, right y-axis), plotted for the period till June 1, 2020. Zooming in at the beginning of the pandemic allows us to see the most interesting period in the epidemic spread.

Figure 2 compares difference in $R_t$ curve shapes for different countries in more visual way, by placing them onto one graph. In the beginning of the outbreak the US had the highest values of reproduction number and Japan - the lowest one; at the end of the year the highest $R_t$ was observed for Sweden.

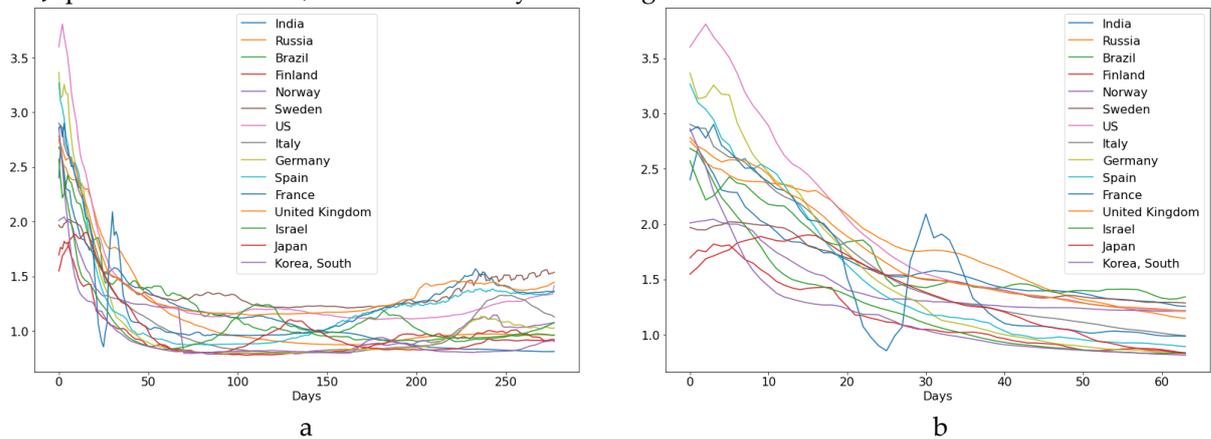

**Figure 2.** Comparison of $R_t$ shapes for different countries, on the x-axis the number of days since the first 1000 infected cases in each country. a) Left graph shows $R_t$ shapes for dates till December 31, 2020, b) right graph – till June 1, 2020

### 3.2. Relation between $R_t$ and peoples mobility

One of our main goals was to study the connection between different measures to limit people's mobility and the spread of the epidemic, measured by $R_t$. To estimate people's mobility we use Apple's mobility index. Visual relationships between $R_t$ and the mobility in different countries are presented in Figure 3a.

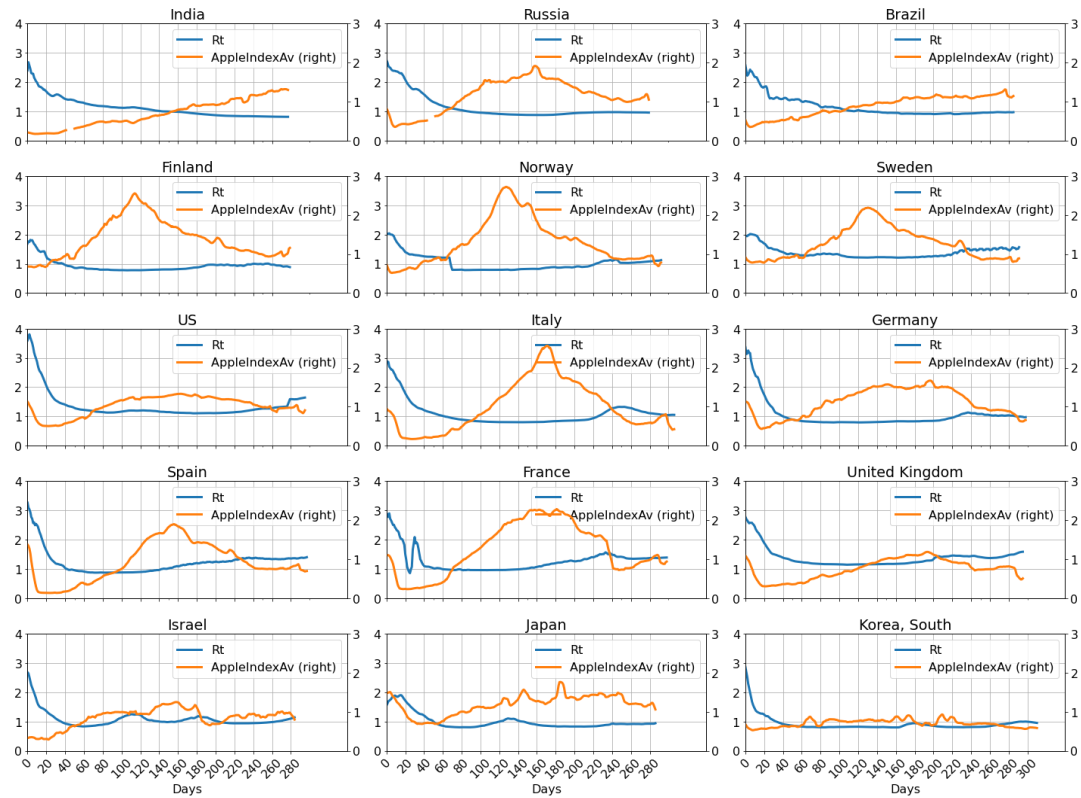

**Figure 3a.** Apple mobility index (orange line, right y-axis) and time-dependent reproduction number $R_t$ (blue line, left y-axis), plotted for the period till December 31, 2020. Time-dependent reproduction number $R_t$ and Apple mobility index (aggregated as mean of three data streams: 'driving', 'walking' and 'transit' mobility) smoothed by 7-days sliding window. On the x-axis the number of days since the first 1000 infected cases in each country till December 31, 2020 is shown; x-axis and y-axis are shown in the same scale for all countries.

There are a few conclusions that can be derived from observing those figures:

- If we consider the long time period (till December 31, 2020), over time there seems to be relatively low correlation between mobility constraints and $R_t$, similarly to the results found in [22].
- For some countries (India and Brazil), Apple mobility index shows steady growth, and does not seem to depend on any mobility prevention measures introduced by local government.
- In many cases, especially considering some local time intervals, there seems to be negative correlation between $R_t$ and mobility index. This can be explained by the fact that preventive measures are introduced as a result of rising $R_t$, and thus when $R_t$ is still rising, preventive measures are already in place, affecting mobility index.

The relationship between $R_t$ and mobility index for the first part of the pandemic is presented in Figure 3b. On this figure, for many countries one can see the initial drop of both $R_t$ and the mobility index, corresponding to the initial lockdown, which is followed by the rise of mobility index (removal of lockdown measures).

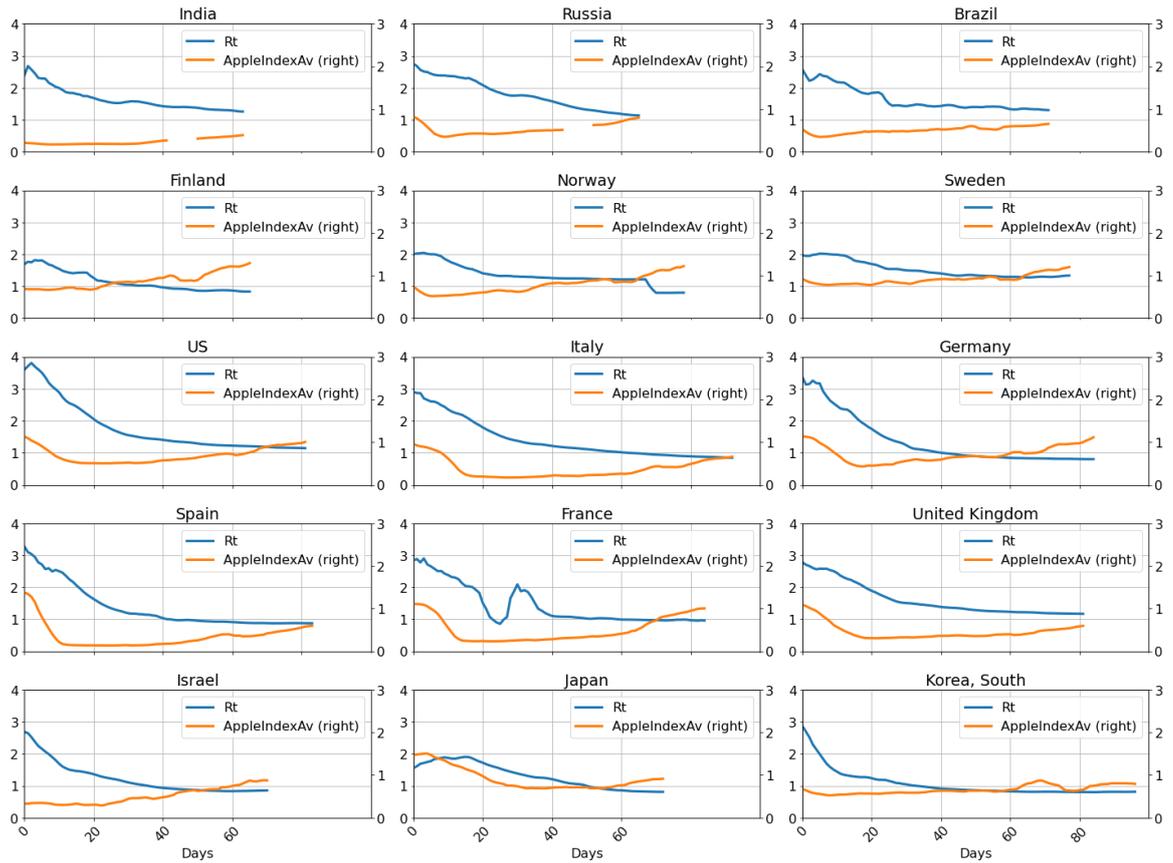

**Figure 3b.** Apple mobility index (orange line, right y-axis) and time-dependent reproduction number $R_t$ (blue line, left y-axis), plotted for the period till June 1, 2020. For many countries one can see the initial drop of both $R_t$ and mobility index, corresponding to the initial lockdown, which is followed by the rise of mobility index (removal of lockdown measures).

Our results support the reasoning proposed in [22], which indicates that protective measures become less effective in the middle of the pandemic, but can have stronger effect in the beginning.

*3.2. Relation between derivative of $R_t$ and people's mobility*

From the Figure 3a it follows, that there is a strong negative correlation between $R_t$ and mobility index for some time intervals. This may be caused by the fact that lockdown measures are implemented as a result of rising $R_t$, and their effect on $R_t$ is not immediate. Our hypothesis, however, is that the rate of increase of $R_t$ starts to drop immediatly as a result of lockdown measures. Similarly, when mobility measures are loosened, the change rate of $R_t$ ($dR_t/dt$) increases.

For some countries we can suggest the significant correlation between $dR_t/dt$ and mobility index for the very beginning of the pandemic, while for all 2020 for most of the countries the significant correlation is not observed (Figure 4a). Only for some countries like Israel, Japan and South Korea we can suggest the significant correlation on the whole time period, and those are the countries with the most strictly imposed restrictions (another such country would be China, but the mobility index data is not available).

The non-monotonous curve of $R_t$ change rate in combination with monotonous mobility curve may suggest that the reproduction number is influenced also by non-mobility events (for example, massive violations of social distance or masks wearing). Local extrema of the $R_t$ derivative curve shows the suggested dates of such events.

Figure 4b zooms in to show the relationship between $dR_t/dt$ and mobility index for the first phase of the pandemic, until June 1, 2020.

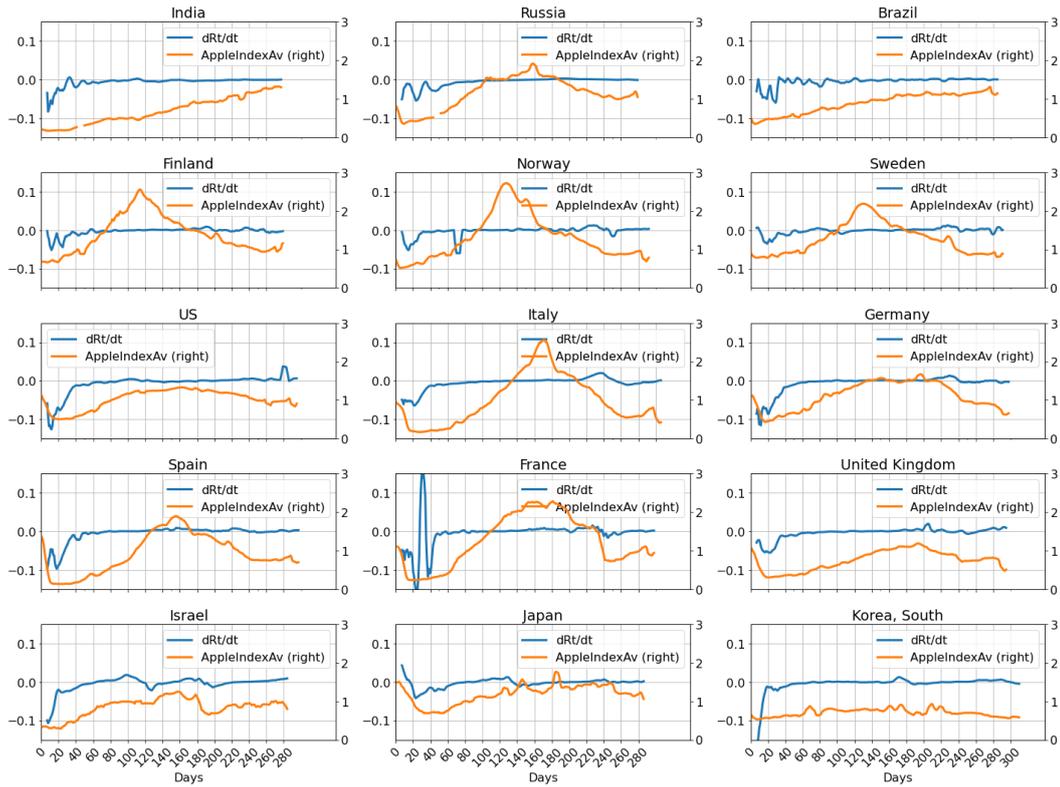

**Figure 4a.** Apple mobility index (orange line, right y-axis) and $R_t$ change rate (blue line, left y-axis), plotted for the period till December 31, 2020. The $R_t$ change rate and smoothed Apple mobility index (aggregated as a mean of three data streams: 'driving', 'walking' and 'transit' mobility). On the x-axis the number of days since the first 1000 infected cases in each country till December 31, 2020 is shown; x-axis and y-axis are shown in the same scale for all countries.

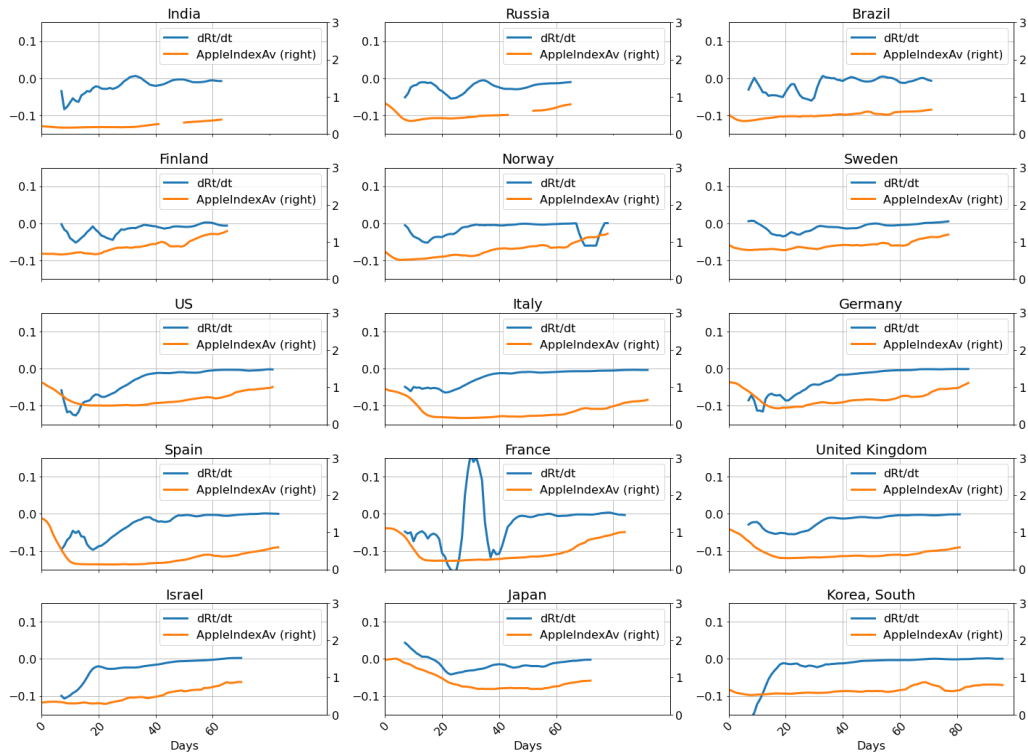

**Figure 4b.** Relationsip between Apple mobility index (orange line, right y-axis) and $R_t$ change rate (blue line, left y-axis) for the first phase of the pandemic, plotted for the period till June 1, 2020.

*3.3. Further look at the correlation*

To understand the association between observed values of $R_t$, $R_t$ change rate ($dR_t/dt$) and mobility we showed the Spearman correlation matrix for different countries (Figure 5). The spread of COVID-19 in 2020 showed a negative correlation of $R_t$ for most of the countries (Israel, Spain and Japan have correlation cofficients close to zero). The correlation of $R_t$ change rate with people's mobility is positive for all of the countries and strong for Italy, Germany and Spain.

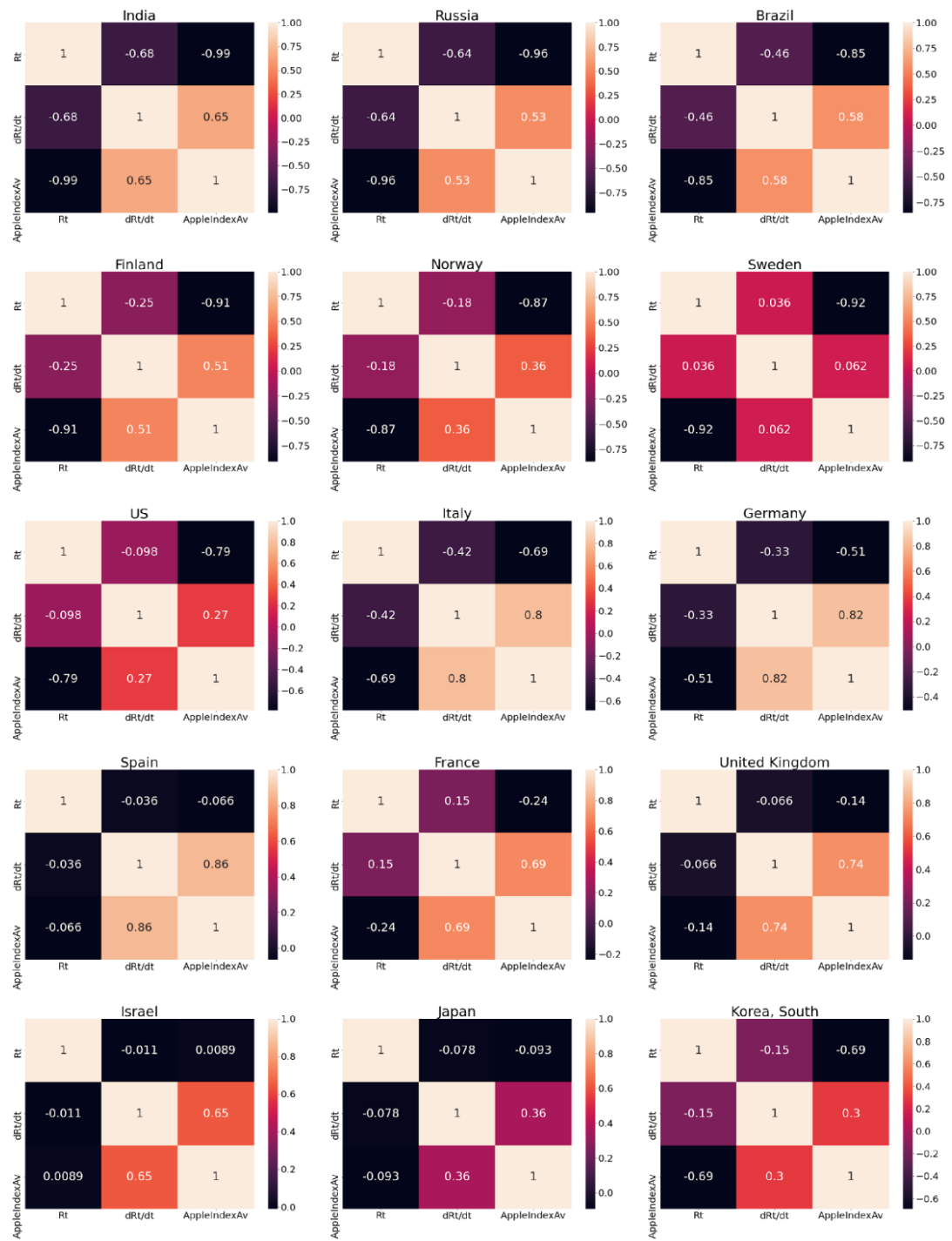

**Figure 5.** Spearman correlation matrix correlation for Apple mobility index, $R_t$, $R_t$ change rate ($dR_t/dt$) in different countries for 2020, plotted for the period till December 31, 2020. The correlation of $R_t$ change rate ($dR_t/dt$) and people's mobility is positive for all of the countries and strong for Italy, Germany and Spain.

Figure 5 shows the general trend over the whole year, and suggest the following conclusions:

- Introduction of mobility restrictions lead to change of $dR_t/dt$, and vice versa, once the restrictions are removed – it leads to rise of the mobility index and increase of $dR_t/dt$.
- Most countries introduce mobility restrictions as a response to epidemic spread, characterized by rising $R_t$ values. There is significant delay between introduction of mobility restrictions and changes in direction of $R_t$, which leads to anti-correlation.

*3.4. Investigating the delay of effectiveness of mobility prevention measures*

The latter point suggests that we need to account for some delay between the time that mobility restrictions are introduced, and the time they start having an effect on $R_t$ changing its direction. To find this delay, we computed Spearman correlation between $R_t$ and mobility index with some time shift. The graphs showing the relationship between the time shift and Spearman correlation for different countries is shown in Figure 6. A shift of 5 days means that the mobility index is shifted by 5 days from $R_t$.

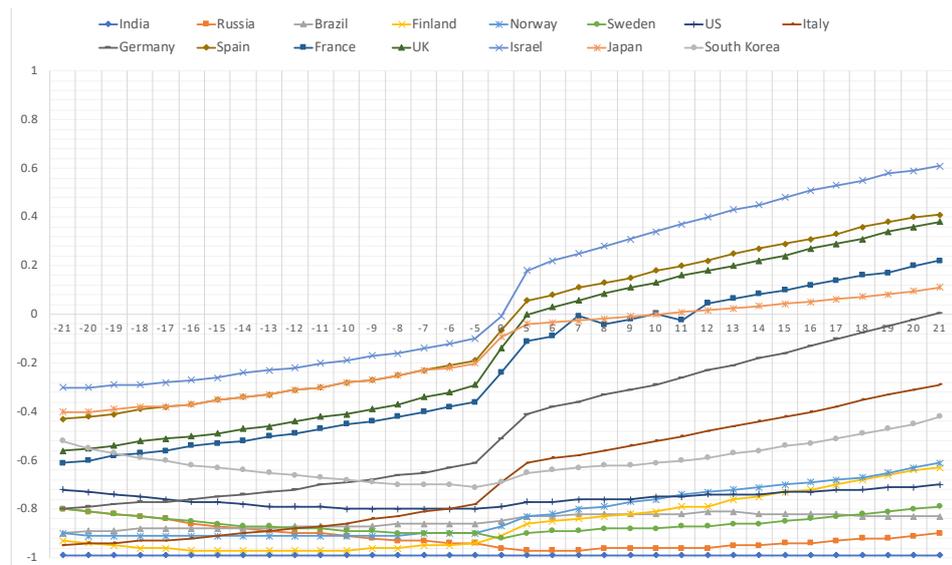

**Figure 6.** Spearman correlation for $R_t$ and mobility for different delay intervals. Sensitivity analysis for different countries was performed with varying delay intervals (5 or 21 days) between mobility and $R_t$ to account for delay in measures efficiency.

- When the time shift is moved to more negative values area, for most countries, an increase in the correlation modulus is observed, which corresponds to a decrease in its absolute value. This may indicate a delay in the introduction of restrictive measures when the epidemic spreads.
- When the time shift is moved to more positive values, for most countries, an increase in the correlation modulus is observed, which corresponds to an increase in its absolute value. This may indicate a delay in the reaction of $R_t$ to the introduced restrictions.

**4. Discussion**

There are some restrictions regarding the applicated model. For $R_t$ calculation we do not separate internal and external infected cases, assuming that all incident cases after the first time-point arise from local transmission, i.e. it does not account for the possibility that cases (other than those appearing at the first timestep) are imported from other locations or derived from alternative host species. This assumption is used, for example, by [34]. In our case, it is applicable since we try to calculate the potential velocity of epidemic spreading and the way infection appears is not a case of the study. Based on previous studies, we also suspect that for the first year of epidemic spread a recovered person cannot become susceptible again [35].

Our model doesn't contain any distributions so we showed only the baseline of $R_t$, without any confidence interval. It should also be mentioned that Apple mobility data is only generated by Apple users with location services enabled. In some countries these people may be a minority in the population.

If we consider the long time period (till December 31, 2020), over time there seems to be relatively low correlation between mobility constraints and $R_t$. These findings support the reasoning proposed in [22], which indicates that protective measures become less effective in the middle of the pandemic, but can have stronger effect in the beginning.

## 5. Conclusions

The developed open-source tool for estimating reproduction number based on SEIR model in real-time allows to forecast the progression of disease outbreaks at any phase of epidemic and understand the effect of interventions on epidemics spread. This tool also allows to assess the disease transmission potential and can be easily adapted to future outbreaks of different pathogens. The tool calculates $R_t$ based on the fitting of SEIR model predictions to actual values, and thus is much less susceptible to data noise, like the traditional way of $R_t$ estimation.

For COVID-19 spread in 2020 we showed that for most of the countries the correlations between the $R_t$ and people's mobility (based on the Apple mobility index) are negative. This shows a delay in the implementation of mobility restrictions – the countries imposed them in response to growth of new COVID-19 cases, rather than preventively.

**Supplementary Materials**: A tool for estimating the time-dependent reproduction number during SARS-COVID-19 pandemic for countries worldwide is available online as Python code from the following Github repository: https://github.com/shwars/COVID19Modelling

**Author Contributions:**

Conceptualization, T.P., D..S. and A.G.; methodology, T.P. and A.G.; data analysis, visualization and validation, T.P. and D.S.; writing—original draft preparation, T.P.; writing—review and editing, T.P., D..S. and A.G. The authors contributed equally to this work. All authors have read and agreed to the published version of the manuscript.

**Acknowledgments:** This work was supported as part of the research program of MSU Center for Storage and Analysis of Big Data.

**Data Availability Statement:** All data used are publicly available.

**Conflicts of Interest:** The authors declare no conflict of interest.